# Discovery of New Layered Iron Arsenide Superconductor (Ca,Pr)FeAs$_2$

Hiroyuki Yakita[*,1], Hiraku Ogino[1], Tomoyuki Okada[1], Akiyasu Yamamoto[1], Kohji Kishio[1], Tetsuya Tohei[2], Yuichi Ikuhara[2], Yoshito Gotoh[3], Hiroshi Fujihisa[3], Kunimitsu Kataoka[3], Hiroshi Eisaki[3], Jun-ichi Shimoyama[1]

[1]Department of Applied Chemistry, The University of Tokyo, 7-3-1 Hongo, Bunkyo, Tokyo 113-8656, Japan
[2]Institute of Engineering Innovation, The University of Tokyo, 2-11-16 Yayoi, Bunkyo, Tokyo 113-8656, Japan
[3]National Institute of Advanced Industrial Science and Technology (AIST), Tsukuba, Ibaraki 305-8565, Japan

**ABSTRACT:** A new iron-based superconductor (Ca,Pr)FeAs$_2$ was discovered. Plate-like crystals of the new phase were obtained and crystal structure was investigated by single-crystal X-ray diffraction analysis. The structure was identified as the monoclinic system with space group $P2_1/m$, and is composed of two Ca(Pr) planes, anti-fluorite Fe$_2$As$_2$ layers, and As$_2$ zigzag chain layers. Plate-like crystals composed of the new phase showed superconductivity with $T_c$ ~20 K in both magnetization and resistivity measurements.

Several groups of iron-based superconductors, such as $RE$FeAs(O,F)[1] ($RE$ = rare earth elements), $AE$FeAsF[2] ($AE$ = alkaline earth metals), $AE$Fe$_2$As$_2$[3], LiFeAs[4], FeSe[5], and compounds having perovskite-type oxide layer (ex. Fe$_2$P$_2$Sr$_4$Sc$_2$O$_6$[6]) have been discovered since 2008. Developments of superconducting tapes and wires have been already attempted for potassium doped $AE$Fe$_2$As$_2$ and fluorine doped $RE$FeAsO, because of their high $T_c$ and high $H_{c2}$. On the other hand, discovery of new superconductor with high $T_c$ and high chemical stability has been still desirable. Since iron-based superconductors are composed of stacking of superconducting layers of Fe$_2$Pn$_2$ or Fe$_2$Ch$_2$) ($Pn$ = P, As, $Ch$ = S, Se, Te) and blocking layers, design and search of new blocking layers are promising for discovery of new superconductors. There are several compounds having anti-fluorite blocks and As-based blocks such as UCuAs$_2$[7]. Recently new iron based superconductors Ca$_{10}$(Pt$_3$As$_8$)(Fe$_2$As$_2$)$_5$ and Ca$_{10}$(Pt$_4$As$_8$)(Fe$_2$As$_2$)$_5$[8-10] were reported. These compounds have As–based blocking layers between Fe$_2$As$_2$ layers. The structures of these compounds indicate that there are possibilities to find new iron-based superconductors with arsenide blocking layers. In addition, Saha et al. reported that single crystalline (Ca,$RE$)Fe$_2$As$_2$ ($RE$ = La, Ce, Pr, Nd) showed superconductivity with high $T_c$ exceeding 40 K in resistivity measurement[11]. After this report, superconductivity in $RE$-doped CaFe$_2$As$_2$ has been reported by several groups[12]. On the other hand, there are several works that report coexistence of two superconducting phases or interface superconductivity in this system[13,14]. Therefore, new superconducting phase is expected in Ca-$RE$-Fe-As system.

In the present study, we have explored new iron-based superconductors in Ca-Pr-Fe-As system and found a new compound (Ca,Pr)FeAs$_2$. We report a crystal structure and physical properties of (Ca,Pr)FeAs$_2$.

All samples were synthesized by the solid-state reaction starting from FeAs(3N), PrAs(3N), Ca(2N), As(4N), and CaO(3N). Since the starting reagents, PrAs and Ca are sensitive to moisture and/or oxygen in air, manipulations were carried out in an argon-filled glove box. Powder mixtures were pelletized, sealed in evacuated quartz ampoules, and reacted at 1000~1200°C for 24 h. Constituent phases were studied by the powder XRD measurements using RIGAKU Ultima-IV diffractometer and intensity date were collected in the $2\theta$ range of 5°-80° at a step of 0.02° using Cu-$K_\alpha$ radiation. Local chemical composition of the samples was investigated by a scanning electron microscope (KEYENCE VE-7800) equipped with an energy dispersive X-ray spectrometer (EDX). Magnetic susceptibility was evaluated by a SQUID magnetometer (Quantum Design MPMS-XL5s). Electrical resistivity was measured by the AC four-point-probe method using Physical Property Measurement System (Quantum Design). High-angle annular dark field (HAADF) images of the samples were obtained using a scanning transmission electron microscope (STEM: JEM-ARM200F, JEOL). In the structure analysis of (Ca,Pr)FeAs$_2$, integrated intensity data were collected by single-crystal X-ray diffractometer with an imaging plate (Rigaku R-AXIS RAPID-II) using graphite-monochromatized Mo $K_\alpha$ radiation ($\lambda$ = 0.71069 Å) at room temperature. In our data collection, 2518 reflections were measured and 402 unique reflections with $|F_{obs.}| > 3\sigma(|F_{obs.}|)$ were considered to be observed ones. Lorentz polarization corrections and absorption corrections were applied to all of the collected reflections. All of the calculations for the structure analysis of (Ca,Pr)FeAs$_2$ were carried out using the computer program Superflip[15] and FMLSM[16].

Figure 1 shows powder XRD patterns of (Ca$_{0.9}$Pr$_{0.1}$)Fe$_{1.9}$As$_{2.1}$, (Ca$_{0.9}$Pr$_{0.1}$)Fe$_{1.3}$As$_{1.8}$O$_{0.2}$ and (Ca$_{0.9}$Pr$_{0.1}$)FeAs$_2$ sintered at 1000~1200°C. Diffraction peaks due to FeAs were observed in all samples. In addition, peaks of CaFe$_2$As$_2$ and PrAs appeared in (Ca$_{0.9}$Pr$_{0.1}$)Fe$_{1.9}$As$_{2.1}$ and $(Ca_{0.9}Pr_{0.1})Fe_{1.3}As_{1.8}O_{0.2}$ samples, and $FeAs_2$ was formed in $(Ca_{0.9}Pr_{0.1})FeAs_2$ sample. Samples of $(Ca_{0.9}Pr_{0.1})Fe_{1.3}As_{1.8}O_{0.2}$ and $(Ca_{0.9}Pr_{0.1})FeAs_2$ showed several unidentified peaks marked by stars in Figure 1, suggesting that the generation of a new layered compound with a layer spacing of 10.4 Å. Note that the peaks appeared only in the samples starting from iron-poor composition relative to arsenic. Furthermore, we confirmed that Pr-doping was also needed to form the present new phase. Plate-like crystals which are composed of the new phase was successfully separated from a sintered bulk sample with a nominal composition of $(Ca_{0.9}Pr_{0.1})Fe_{1.3}As_{1.8}O_{0.2}$. The surface XRD patterns and an optical microscope image of a plate-like crystal are shown in Figure 2. Edges of the crystals were slightly faceted as shown in the inset. All sharp peaks can be assigned as $00l$ reflections when the layered crystal structure with interlayer distance of 10.4 Å is assumed. Compositional analysis carried out for such crystals using EDX indicated that their atomic ratio is (Ca,Pr) : Fe : As = 24.1 : 23.5 : 52.4, which is close to (Ca,Pr) : Fe : As = 1 : 1 : 2. Concentration of Pr was about 17 % of that of Ca site. Addition of CaO promoted formation of the plate-like crystals, though oxygen could not be detected in the crystals by our EDX.

A single crystal of the new phase with dimensions approximately 0.15 x 0.12 x 0.005 $mm^3$ was used for the structural analysis. The structural analysis revealed that chemical formula of the new phase was $(Ca_{1-x}Pr_x)FeAs_2$, corresponding to the result of compositional analysis of EDX. Figure 3 shows the crystal structure of $(Ca_{1-x}Pr_x)FeAs_2$. The lattice constants of monoclinic structure of $(Ca_{1-x}Pr_x)FeAs_2$ were refined as $a$ = 3.9163(8), $b$ = 3.8953(7), $c$ = 10.311(3) Å and $\beta$ = 90.788(8)°. We have observed a reflection condition for $0k0$ data as $k = 2n$. This leads to the centrosymmetric space group $P2_1/m$ for $(Ca_{1-x}Pr_x)FeAs_2$. The structure of $(Ca_{1-x}Pr_x)FeAs_2$ was refined with 402 unique data. In the final refinement, $(\Delta/\sigma)_{max.} < 0.01$ was fully satisfied, where $\Delta$ is the shift of parameters and $\sigma$ is the standard uncertainty. The final atomic coordinates and equivalent isotropic atomic displacement parameters are given in Table 1, where the $R$ value converged to 0.116 and the $Rw$ ($w = 1/\sigma^2(F_{obs})$) value was 0.157. Concentration of Pr obtained by the structure analysis was 27 % of the calcium site. Although this value is larger than that estimated by the EDX analysis, both results suggested the new phase has higher Pr occupancy at the Ca site than the nominal ratio.

Figure 4 (a) and (b) show HAADF-STEM images of the plate-like sample with a nominal composition of $(Ca_{0.9}Pr_{0.1})Fe_{1.3}As_{1.8}O_{0.2}$ taken from [110] and [010] direction, respectively. As clearly seen, the observed crystal has a layered structure with an interlayer distance of approximately 10.4 Å, which corresponds well with the structure analysis. A stacking of anti-fluorite type $Fe_2As_2$ layer, Ca(Pr) planes and $As_2$ chain layers were clearly observed.

Figure 5 shows the temperature dependences of zero-field cooled (ZFC) and field-cooled (FC) magnetization curves of polycrystalline sample of $(Ca_{0.9}Pr_{0.1})Fe_{1.9}As_{2.1}$ and powder sample of $(Ca_{0.9}Pr_{0.1})FeAs_2$, and about ten plate-like crystals obtained from a sample with a nominal composition of $(Ca_{0.9}Pr_{0.1})Fe_{1.3}As_{1.8}O_{0.2}$. The typical size of crystals was ~0.5 x 0.5 x 0.01 $mm^3$ and magnetic field was applied normal to their wide surface. Since the sintered bulks of $(Ca_{0.9}Pr_{0.1})FeAs_2$ were easy to deteriorate in air, its magnetization behavior was measured for powder sample. Diamagnetism was not observed in a polycrystalline sample of $(Ca_{0.9}Pr_{0.1})Fe_{1.9}As_{2.1}$, while large diamagnetism suggesting bulk superconductivity were observed in the samples containing the new phase, $(Ca_{0.9}Pr_{0.1})Fe_{1.3}As_{1.8}O_{0.2}$, with $T_{c(onset)}$ of ~20 K. For $(Ca_{0.9}Pr_{0.1})FeAs_2$ powder, relatively small diamagnetism does not deny its bulk superconductivity, because volume fraction of the new phase was small as shown in Figure 1 and the grain size of the powder sample is comparable to the penetration depth.

Figure 6 shows temperature dependence of resistivity of a small plate-shaped bulk which consist of plate-like crystals and was separated from a sintered bulk of $(Ca_{0.9}Pr_{0.1})Fe_{1.3}As_{1.8}O_{0.2}$. In this measurement, AC current was applied parallel to the wide plane of the sample and magnetic field was applied normal to the wide surface. Those experimental conditions mean that AC current and magnetic field were applied parallel and normal to the iron-arsenide layer, respectively, of the crystals of the new phase. Superconducting transition was also confirmed in the resistivity measurement with $T_{c(onset)}$ of ~24 K. However, the transition was relatively dull and zero resistance was achieved at ~12 K under zero external field. Since the current density in this measurement was less than 0.1 $A/cm^2$, the broad transition is probably due to distribution of $T_c$ in the sample reflecting inhomogeneous praseodymium concentration and/or weak coupling between plate-like crystals in a small bulk. Although slight increases in resistivity by applied magnetic field can be confirmed up to ~40 K, we believe this is not an intrinsic behavior of the new phase. This sample may contain very small amount of 40 K-class superconducting regions.

$(Ca,Pr)FeAs_2$ did not show 40 K-class high $T_c$ as observed in $(Ca,Pr)Fe_2As_2$[11, 12]. In $Ca_{0.73}Pr_{0.27}FeAs_2$, As-Fe-As bond angle lies between 106.2~111.2° while $CaFe_2As_2$, composed of alternate stacking of Ca planes and $Fe_2As_2$ layers, has almost ideal As-Fe-As bond angle. We tentatively attribute 20 K class $T_c$ of $(Ca,Pr)FeAs_2$ to its As-Fe-As bond angles, because they are far from ideal value of 109.47°. In As chain layers, the closest As-As distance is about 2.60 Å, which is similar to As-As covalent bond distance (2.42 Å)[17] as well as that of $[As_2^{4-}]$ dimers (2.50 Å) in $Ca_{10}(Pt_4As_8)(Fe_2As_2)_5$[9], suggesting existence of two As-As bonding for each As atom in this layer. In addition, zigzag chain structure of pnictide is reported in several compounds such as KAs[18] and $SrZnSb_2$[19]. Considering these factors, the valence of As atom in As chain layer seems to be -1, meaning $(Ca,Pr)FeAs_2$ is composed of stacking of $Fe_2As_2^{2-}$ layer, two sheets of $Ca^{2+}$ plane, and $As_2^{2-}$ chain layer. Thus, $CaFeAs_2$ can be viewed as the parent compound of new series of iron-based superconductors, and Pr substitution to Ca site works as electron doping. At present we could not succeed to control Pr concentration in the compound, but if it is controlled, phase diagram of this phase will be clarified. Another challenge is that optimization of the crystal structure and enhancement of $T_c$ by changing $RE$ elements and so on.

In summary, a new layered compound $(Ca,Pr)FeAs_2$ has been discovered in Ca-Pr-Fe-As system. Plate-like crystals of the phase were obtained from a sample with the nominal composition of $(Ca_{0.9}Pr_{0.1})Fe_{1.3}As_{1.8}O_{0.2}$ and a new crystal structure composed of $Fe_2As_2$ layer, two sheets of Ca plane, and $As_2$ chain layer was identified by X-ray diffraction analysis. EDX analysis indicated that atomic ratio of the new phase is close to (Ca,Pr) : Fe : As ~1 : 1 : 2. Atomic images of the samples obtained by HAADF-STEM observation were in good agreement with the result of the structure analysis. Samples composed of the new phase exhibited superconductivity with $T_c$ ~20 K in both magnetization and resistivity measurements. Discovery of this new compound proved variety of blocking layers in iron pnictides, and may lead to discoveries of other superconductors.

## AUTHOR INFORMATION

**Corresponding Author**

Hiroyuki Yakita, Department of Applied Chemistry, The University of Tokyo, 7-3-1 Hongo, Bunkyo, Tokyo 113-8656, Japan.
Email: 8757570603@mail.ecc.u-tokyo.ac.jp
Hiraku Ogino, Department of Applied Chemistry, The University of Tokyo, 7-3-1 Hongo, Bunkyo, Tokyo 113-8656, Japan.
Email: tuogino@mail.ecc.u-tokyo.ac.jp

## ACKNOWLEDGMENT

This study was supported by Strategic International Collaborative Research Program (SICORP), Japan Science and Technology Agency, Asahi glass foundation and Tokuyama science foundation, and partly by "Nanotechnology Platform" (Project No. 12024046) of the Ministry of Education, Sports, Science and Technology (MEXT), Japan. We thank Prof. D. Johrendt for valuable discussion on the synthesis of the samples.

*Note added:* Katayama et al., recently reported superconductivity in similar phase.[20]

Table 1. Atomic Coordinates and Equivalent Isotropic Displacement Parameters (Å$^2$) for (Ca$_{0.73}$Pr$_{0.27}$)FeAs$_2$

| atom | Occupancy | x | y | z | 100$U_{eq.}$ (Å$^2$) |
|---|---|---|---|---|---|
| Ca / Pr | 0.730(8) / 0.270 | 0.7432(9) | 0.25 | 0.2297(5) | 1.6(1) |
| Fe | 1 | 0.255(1) | 0.25 | 0.4977(6) | 1.6(2) |
| As(1) | 1 | 0.2460(7) | 0.75 | 0.3598(4) | 1.3(1) |
| As(2) | 1 | 0.2193(8) | 0.25 | 0.0023(4) | 2.3(1) |

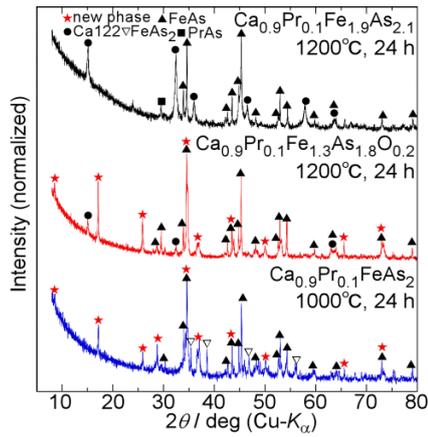

Figure 1. Powder XRD patterns of polycrystalline samples with nominal composition of (C$_{0.9}$Pr$_{0.1}$)Fe$_{1.9}$As$_{2.1}$, (Ca$_{0.9}$Pr$_{0.1}$)Fe$_{1.3}$As$_{1.8}$O$_{0.2}$ and (Ca$_{0.9}$Pr$_{0.1}$)FeAs$_2$.

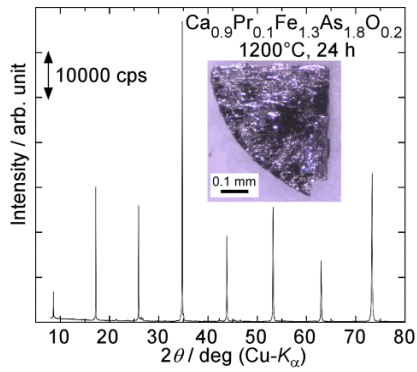

Figure 2. Surface XRD pattern of a plate-like crystal composed of the new phase. Inset shows an optical image of the sample.

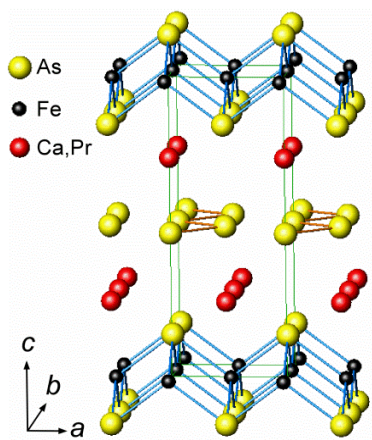

Figure 3. Crystal structure of $(Ca_{0.73}Pr_{0.27})FeAs_2$ with monoclinic structure (space group $P2_1/m$) investigated by single-crystal X-ray structure analysis.

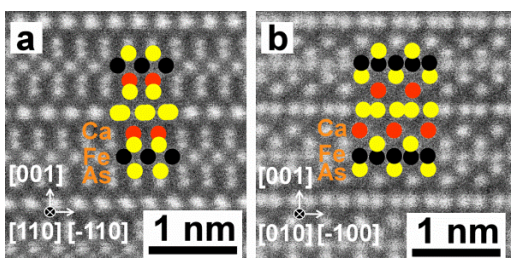

Figure 4. HAADF-STEM image of the plate-like crystal of $(Ca,Pr)FeAs_2$ observed from the [110] direction (a) and [010] direction (b).

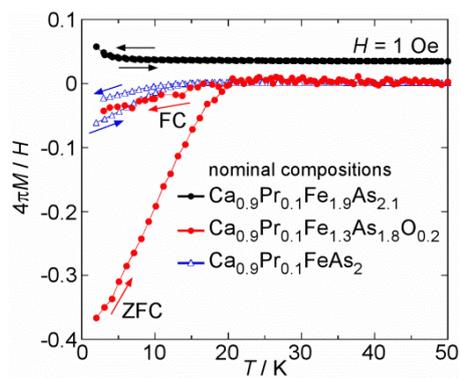

Figure 5. Temperature dependence of ZFC and FC magnetization curves of $(Ca_{0.9}Pr_{0.1})Fe_{1.9}As_{2.1}$, $(Ca_{0.9}Pr_{0.1})Fe_{1.3}As_{1.8}O_{0.2}$ and $(Ca_{0.9}Pr_{0.1})FeAs_2$.

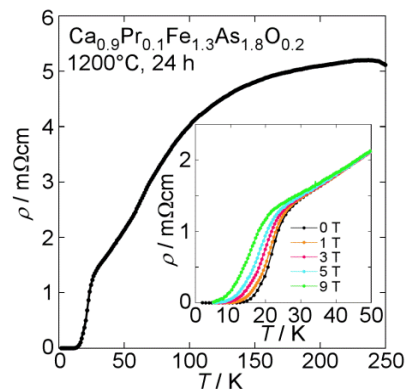

Figure 6. Temperature dependences of resistivity under various magnetic fields of a polycrystalline sample of $(Ca_{0.9}Pr_{0.1})Fe_{1.3}As_{1.8}O_{0.2}$ sintered at 1200°C for 24 h.